\pdfoutput=1
\newcommand*{\ATLASLATEXPATH}{latex/}
\documentclass[atlasdraft=false, texlive=2016, british, cernpreprint]{\ATLASLATEXPATH atlasdoc}
\usepackage{\ATLASLATEXPATH atlaspackage}
\usepackage{multirow}
\usepackage{arydshln}
\usepackage{lscape}
\usepackage{dcolumn}
\usepackage{\ATLASLATEXPATH atlasbiblatex}

\usepackage{\ATLASLATEXPATH atlasphysics}
\graphicspath{{logos/}{figures/}}

\usepackage{SUSY-2016-31-PAPER-defs}

\newcommand{\lumi}{36.1}

\AtlasTitle{Search for heavy charged long-lived particles in proton--proton collisions at $\sqrt{s} = 13$~\TeV\ using an ionisation measurement with the \mbox{ATLAS} detector}

\AtlasVersion{3.0}

\AtlasAbstract{%
This Letter presents a search for heavy charged long-lived particles produced in proton--proton collisions at $\sqrt{s}=13$~\TeV\ at the LHC using a data sample corresponding to an integrated luminosity of $\lumi$~fb$^{-1}$ collected by the \mbox{ATLAS} experiment in 2015 and 2016.
These particles are expected to travel with a velocity significantly below the speed of light, and therefore have a specific ionisation higher than any high-momentum Standard Model particle of unit charge.
The pixel subsystem of the \mbox{ATLAS} detector is used in this search to measure the ionisation energy loss of all reconstructed charged particles which traverse the pixel detector. 
Results are interpreted assuming the pair production of $R$-hadrons as composite colourless states of a long-lived gluino and Standard Model partons. 
No significant deviation from Standard Model background expectations is observed, and lifetime-dependent upper limits on
$R$-hadron production cross-sections and gluino masses are set, assuming the gluino always decays to two quarks and a 100~\GeV\ stable neutralino.
$R$-hadrons with lifetimes above 1.0~ns %with gluino decaying to $q\bar{q}$ plus a 100~\GeV\ neutralino 
are excluded at the $95\%$ confidence level, with lower limits on the gluino mass ranging between 1290~\GeV\ and 2060~\GeV. 
In the case of stable $R$-hadrons, the lower limit on the gluino mass at the $95\%$ confidence level is 1890~\GeV.
}

 \author{The ATLAS Collaboration}

\AtlasRefCode{SUSY-2016-31}

\PreprintIdNumber{CERN-EP-2018-198}
 \arXivId{1808.04095}

 \AtlasJournal{Phys.\ Lett.\ B.}
 \AtlasJournalRef{\PLB 788 (2019) 96}
 \AtlasDOI{https://doi.org/10.1016/J.PHYSLETB.2018.10.055}

 \AtlasCoverSupportingNote{Search for metastable heavy charged particles with large ionisation energy loss in pp collisions at sqrt(s) = 13 TeV using the ATLAS experiment: 2015 and 2016 data}{https://cds.cern.ch/record/2234858/}
 \AtlasCoverCommentsDeadline{18 July 2018}

 \AtlasCoverEgroupEditors{atlas-susy-2016-31-editors@cern.ch}

 \AtlasCoverEgroupEdBoard{atlas-susy-2016-31-editorial-board@cern.ch}

\hypersetup{pdftitle={ATLAS document},pdfauthor={The ATLAS Collaboration}}

\begin{document}

\maketitle

%\tableofcontents

%-------------------------------------------------------------------------------
\section{Introduction}
\label{sec:intro}
%-------------------------------------------------------------------------------

A wide range of physics models that extend the Standard Model (SM) predict the existence of new, massive, long-lived particles (LLPs). These particles appear in proposed solutions to the gauge hierarchy problem~\cite{Susskind:1982mw}, including supersymmetric (SUSY) models that either violate~\cite{Dreiner:1997uz,Berger:2003kc,Barbier:2004ez} or conserve~\cite{Kolda:1997wt,Baer:1998pg,Gates:1999ei,Mafi:1999dg,Chertok:2001ja,Giudice:2004tc,ArkaniHamed:2004yi,Fairbairn:2006gg} $R$-parity. 
$R$-parity is a quantum number defined as ($-$1)$^{3(B-L)+2S}$ where $S$ is the particle spin and $L$ and $B$ are, respectively, its lepton and baryon number.
Within SUSY models, sparticles, including gluinos, may be long-lived, with lifetimes depending, for instance, on the mass hierarchy parameters, or on the size of any $R$-parity-violating coupling~\cite{Randall:2011}. 

The study in this Letter is sensitive to many different models of new physics, in particular those that predict the production of massive particles with 
%\emph{O}(ns) 
lifetimes exceeding 1~ns at LHC energies, such as mini-split SUSY~\cite{Giudice:2004tc,ArkaniHamed:2004fb,Arvanitaki:2013} or anomaly-mediated supersymmetry-breaking (AMSB) models~\cite{ASMB1:Giudice1998,ASMB2:Randall1999}. Results are presented assuming the production of $R$-hadrons as composite colourless states of a gluino together with SM quarks or gluons~\cite{Farrar:1978xj}.
% and the subsequent decay of the gluino to a stable neutralino and a $q\bar{q}$ pair~\cite{Farrar:1978xj}.

Due to their large mass, LLPs are expected to be slow ($\beta\gamma\leq 0.9$\footnote{Here $\beta$ is the speed of the particle relative to the speed of light in vacuum and $\gamma = \frac{1}{\sqrt{1-\beta^2}}$} in a large fraction of cases) and therefore, if charged, to have a specific ionisation larger than any SM particle of unit charge at high momentum. The pixel subsystem~\cite{Aad:2008zz} of the \mbox{ATLAS} detector~\cite{PERF-2007-01} provides measurements of ionisation energy loss ($\dedx$) for charged particles with sufficient accuracy to distinguish such highly ionising particles from SM particles. In this Letter, the $\dedx$ information is used to search for LLPs using a data sample of proton--proton (\emph{pp}) collisions corresponding to an integrated luminosity of $\lumi$~fb$^{-1}$ collected at $\sqrt{s} = 13$~\TeV. This extends the reach beyond that of a previous study~\cite{SUSY-2016-03}, thanks to a tenfold increase of the integrated luminosity and to several improvements to the analysis.
It also extends the reach beyond that of similar studies by CMS~\cite{CMS-EXO-15-010} and ATLAS~\cite{SUSY-2016-02} carried out at the same centre-of-mass energy and dedicated to the search for LLPs 
%which are not expected to decay 
not decaying inside the detector. 

%In this paper, $R$-hadrons with lifetimes such that the majority of their decays occurs inside the ATLAS detector are referred to as metastable; in this case the search strategy focuses on isolated large momentum and large $\dedx$ tracks that do not reach the muon spectrometer. The lifetime coverage extends from around 1 ns to several tens of ns. $R$-hadrons which do not decay within the ATLAS detector are termed stable; the stable search strategy mainly differs from the metastable one in that no muon veto is applied.

%-------------------------------------------------------------------------------
\section{ATLAS detector and ionisation measurement}
\label{sec:detector}
%-------------------------------------------------------------------------------

The ATLAS detector\footnote{ATLAS uses a right-handed coordinate system with its origin at the nominal interaction point (IP) in the centre of the detector and the $z$-axis along the beam pipe. The $x$-axis points from the IP to the centre of the LHC ring, and the $y$-axis points upward. Cylindrical coordinates $(r,\phi)$ are used in the transverse plane, $\phi$ being the azimuthal angle around the $z$-axis. The pseudorapidity is defined in terms of the polar angle $\theta$ as $\eta=-\ln\tan(\theta/2)$, and angular distance is measured in units of $\Delta R \equiv \sqrt{(\Delta \eta)^2 +(\Delta \phi)^2}$.} is a general-purpose detector with a forward--backward symmetric cylindrical symmetry described in detail in Ref.~\cite{PERF-2007-01}.
It consists of a tracker for measuring the trajectories of charged particles inside a 2~T solenoidal magnetic field, followed by calorimeters for measuring the energy of particles that interact electromagnetically or hadronically. 
A muon spectrometer immersed in a toroidal magnetic field surrounds the calorimeters, and provides tracking for muons. 
A two-level trigger system is used to select events~\cite{TRIG-2016-01}. The first-level trigger is implemented in hardware and uses a subset of the detector information. This is followed by the software-based high-level trigger, which runs offline reconstruction and calibration software, reducing the event rate to about 1 kHz.
The detector is hermetic and can therefore measure the magnitude of the missing transverse momentum ($\MET$) associated with each event. 
The tracker is made of three detector systems organised in concentric layers. The outermost layer is made of densely packed proportional gas-filled detectors~\cite{IDET-2015-01}, the radial region from roughly 30~cm to 55~cm is equipped with silicon microstrip detectors~\cite{IDET-2013-01} and the innermost layer is covered by a silicon pixel detector~\cite{Aad:2008zz}, which is described below in some detail as it has a crucial role in this analysis.

The pixel detector typically provides four precision measurements for each track in the region $|\eta| <2.5$ at radial distances of 33~mm, 50~mm, 88~mm and 122~mm from the LHC beam line. The innermost pixel layer is named the insertable B-layer (IBL)~\cite{IBL} and was designed to maintain efficient operation of the pixel system above 2$\times$10$^{34}$~cm$^{-2}$~s$^{-1}$ luminosity, when the next-to-innermost pixel layer begins to lose detection efficiency. 
The hit efficiency of the pixel detector in the data sample used for this analysis still exceeds 99$\%$ in all layers. For each pixel hit the length of time with signal above threshold, known as \emph{time over threshold} (ToT), is digitised and recorded.
The ToT is approximately proportional to the ionisation charge
%~\cite{BaBar_ToT}
 and allows the calculation of the specific ionisation energy, $\dedx$, of a track. The ToT measurement is digitised with four bits in the IBL and eight bits in all other pixel layers. If the dynamic range is exceeded for a particular hit in the IBL an overflow bit is set, while for the other layers the hit is not recorded.

%Relative to the other pixel layers, the IBL has pixels with smaller area and faster electronics, which provide ToT with lower dynamic range and lower accuracy (four bits are used to digitise it compared to eight bits in the other pixel layers). 

%~\cite{Aad:2008zz}. 
%The IBL $\dEdx$ information therefore requires a separate treatment.

The charge released by a moving charged particle is rarely contained within just one pixel; neighbouring pixels registering hits are joined together using a connected component analysis~\cite{paperCCA,PERF-2012-05} to form clusters. The charge of a cluster is calculated by summing the charge of all pixels belonging to the cluster after calibration corrections. %; the IBL overflow information is accounted for separately. 
To avoid loss of charge, only clusters completely contained in sensor fiducial regions are used (e.g.\ clusters cannot be in contact with pixels on the sensor edge). 
The $\dedx$ for each reconstructed track is calculated using the average of the individual cluster ionisation measurements (charge collected in the cluster per unit track length in the sensor), for the clusters associated with a track. To reduce the impact of the tails of the Landau distribution, which is expected to describe the energy deposition distribution, the track $\dedx$ is evaluated using a truncated-mean method. The average is calculated after removing the highest-$\dedx$ cluster, or the two highest-$\dedx$ clusters in the relatively rare case of more than four clusters associated with the track. More details of the calculation of \dedx\ may be found in Ref.~\cite{SUSY-2016-03}.

%-------------------------------------------------------------------------------
\section{Analysis overview}
\label{sec:analysis_overview}
%-------------------------------------------------------------------------------

The search strategy consists of looking for excesses in the mass distribution of reconstructed tracks with high transverse momentum, $\pt$, and large $\dEdx$. The mass value is determined from a parameterisation of the Bethe--Bloch relation and depends on the momentum and \dedx\ of selected tracks. 

%Two signal regions are considered. $R$-hadrons with lifetimes such that the majority of their decays occur inside the ATLAS detector are referred to as metastable; in this case the gluino decays into a stable neutralino and a $q\bar{q}$ pair. The search for metastable $R$-hadrons focuses on isolated high-momentum and high-$\dedx$ tracks from charged particles that do not reach the muon spectrometer. The metastable $R$-hadron lifetime coverage extends from around 1~ns to several tens of ns. $R$-hadrons which do not decay within the ATLAS detector are termed stable; the search for stable $R$-hadrons mainly differs from the metastable one in that no muon veto is applied, as some of the $R$-hadrons that pass through the muon spectrometer are reconstructed as muons. 
Two signal regions are considered, and the selection is detailed in Section~\ref{sec:selection}. The first region targets metastable $R$-hadrons with lifetimes such that the majority of their decays occur inside the detector. In this region, charged particles that reach the muon spectrometer are removed and the selections are optimised for $R$-hadrons with lifetimes from around 1~ns to several tens of ns. A second signal region targets stable $R$-hadrons which do not decay within the detector. In this region, no muon veto is applied, since some of the stable $R$-hadrons that pass through the muon spectrometer are reconstructed as muons.

Events are selected using the lowest-threshold unprescaled calorimetric \MET trigger. In metastable $R$-hadron events, the measured $\MET$ largely originates from neutralinos which carry away unmeasured momenta. In stable $R$-hadron events, the $R$-hadrons leave only modest energy depositions in the calorimeters~\cite{Mackeprang:2006gx} and only a fraction are reconstructed as muons due to their late arrival time in the muon spectrometer. Therefore, most of the momenta of $R$-hadrons are not accounted for in the measurement of \MET, and only QCD initial-state radiation (ISR) provides a visible contribution that results in a measured imbalance. Due to the neutralinos, the $\MET$ trigger efficiency is higher for metastable than for stable $R$-hadrons. The track reconstruction efficiency is, on the contrary, higher for the stable $R$-hadrons and penalises particles with lifetimes shorter than 10~ns, which may not have crossed enough detector layers. The searches for stable and metastable $R$-hadrons require slightly different optimisations.

%The $\MET$ trigger efficiency decreases with the $R$-hadron lifetime while the track reconstruction efficiency increases with the $R$-hadron lifetime. The search for stable and metastable $R$-hadrons requires therefore a slightly different optimisation. The difference between the stable and the metastable selection is detailed in  Section~\ref{sec:selection}; it depends on the track isolation and its matching to a muon.

The background is estimated with a data-driven approach, as described in Section~\ref{sec:background}. Data control samples are used to parameterise the momentum and $\dEdx$ distributions and their interdependence, and then to generate pseudo data which predicts the background distribution. The potential signal contamination is minimised in these background samples by inverting some of the selection criteria.

%-------------------------------------------------------------------------------
\section{Data and simulation}
\label{sec:samples}
%-------------------------------------------------------------------------------

This search uses data from $pp$ collisions at $\sqrt{s} = 13$~\TeV\ provided by the LHC in 2015 and 2016. The integrated luminosity of the data sample is $\lumi$~fb$^{-1}$, after requirements on detector status and data quality have been applied. Further detector-level cleaning selections are applied to the data to reject events affected by calorimeter noise and data corruption.

An additional data sample, collected in a dedicated low-luminosity run in 2016, is used for the calibration of \dedx\ and mass; it consists of randomly triggered events in bunch crossings where collisions are expected and amounts to about 0.4~nb$^{-1}$.

Simulation samples are used to determine the efficiency and associated uncertainty for selecting signal events. To model signal events, the pair production of gluinos with masses between 400~\GeV\ and 3000~\GeV\ was simulated in \textsc{Pythia 6.4.27}~\cite{Sjostrand:2006za} at leading order with the AUET2B~\cite{ATL-PHYS-PUB-2011-014} set of tuned parameters for the underlying event and the CTEQ6L1~\cite{ATLAS_PDF_set} parton distribution function (PDF) set. Dedicated routines~\cite{Kraan:2004tz} were used to hadronise the gluinos; after hadronisation, about 2/3 of the events contain at least one charged $R$-hadron. All sparticles except the gluino and the lightest neutralino are decoupled. The Monte Carlo (MC) signal samples include a modelling of pile-up, adding the expected number of minimum-bias $pp$ interactions from the same and nearby bunch crossings.

In order to more accurately model ISR in the signal events, additional samples of gluinos were generated at leading order with up to two additional partons using \textsc{MadGraph5}{\_a}\textsc{MC@NLO}~\cite{Madgraph},  
interfaced to the \textsc{Pythia 8.186}~\cite{Sjostrand:2007gs} parton shower model. The NNPDF2.3LO~\cite{Ball:2012cx} PDF set is used along with the A14~\cite{ATL-PHYS-PUB-2014-021} set of tuned parameters. The distribution of the transverse momentum of the gluino--gluino system simulated with \textsc{Pythia 6.4.27} was reweighted to match that obtained in the samples simulated with \textsc{MadGraph5}{\_a}\textsc{MC@NLO}.

Simulated events undergo full detector simulation~\cite{ATL_sim} based on a \textsc{Geant4}~\cite{Agostinelli:2002hh} framework; the hadronic interactions of $R$-hadrons with the detector were handled by dedicated \textsc{Geant4} routines based on the model described in Refs.~\cite{Kraan:2004tz,Mackeprang:2006gx,Mackeprang:2009ad}. Signal samples were generated both for non-decaying gluinos, and for gluinos with a set of lifetimes ranging from 1.0~ns to 50~ns which decay into SM quarks and a 100~\GeV\ stable neutralino via the process $\tilde{g}\to q\bar{q} \tilde{\chi}_1^0$. The decay of the $R$-hadrons and the fragmentation and hadronisation of the resulting quarks were performed with a modified version of \textsc{Pythia 6.4.27}. 

To normalise the number of expected signal events, gluino pair production cross-sections are calculated at next-to-leading order in the strong coupling constant, including the resummation of soft-gluino emissions at next-to-leading-logarithm accuracy~\cite{Beenakker:1996ch,Kulesza:2008jb,Kulesza:2009kq,Beenakker:2009ha,Beenakker:2011fu}. The nominal cross-section values and uncertainties are taken from an envelope of cross-section predictions using different PDF sets and factorisation and renormalisation scales, as described in Ref.~\cite{Borschensky:2014cia}. 

%The characteristic signature of this analysis is an event with moderate \MET and at least one track which travels through at least seven layers of the silicon detector, with energy deposition in the pixel detector consistent nearly twice the ionisation energy of a MIP, and momentum greater than 150 GeV. This signature is sensitive to many different massive, charged particles with lifetimes \emph{O}(ns) or longer predicted by beyond-the-SM theories. The efficiency of the ionisation and momentum measurements are largely model-independent as long as they are above threshold; the main model dependencies arise from the lifetime of the long-lived particle, which affects the probability of the particle travelling at least $37$~cm,  and from the kinematics of the long-lived particles' decay products, which affects the efficiency of the \MET trigger and offline selection. 

%To do: double check references for scattering

%-------------------------------------------------------------------------------
\section{$\dEdx$ corrections and mass calculation}
\label{sec:mass}
%-------------------------------------------------------------------------------

\begin{figure}[]
\resizebox{0.65\textwidth}{!}{
\includegraphics{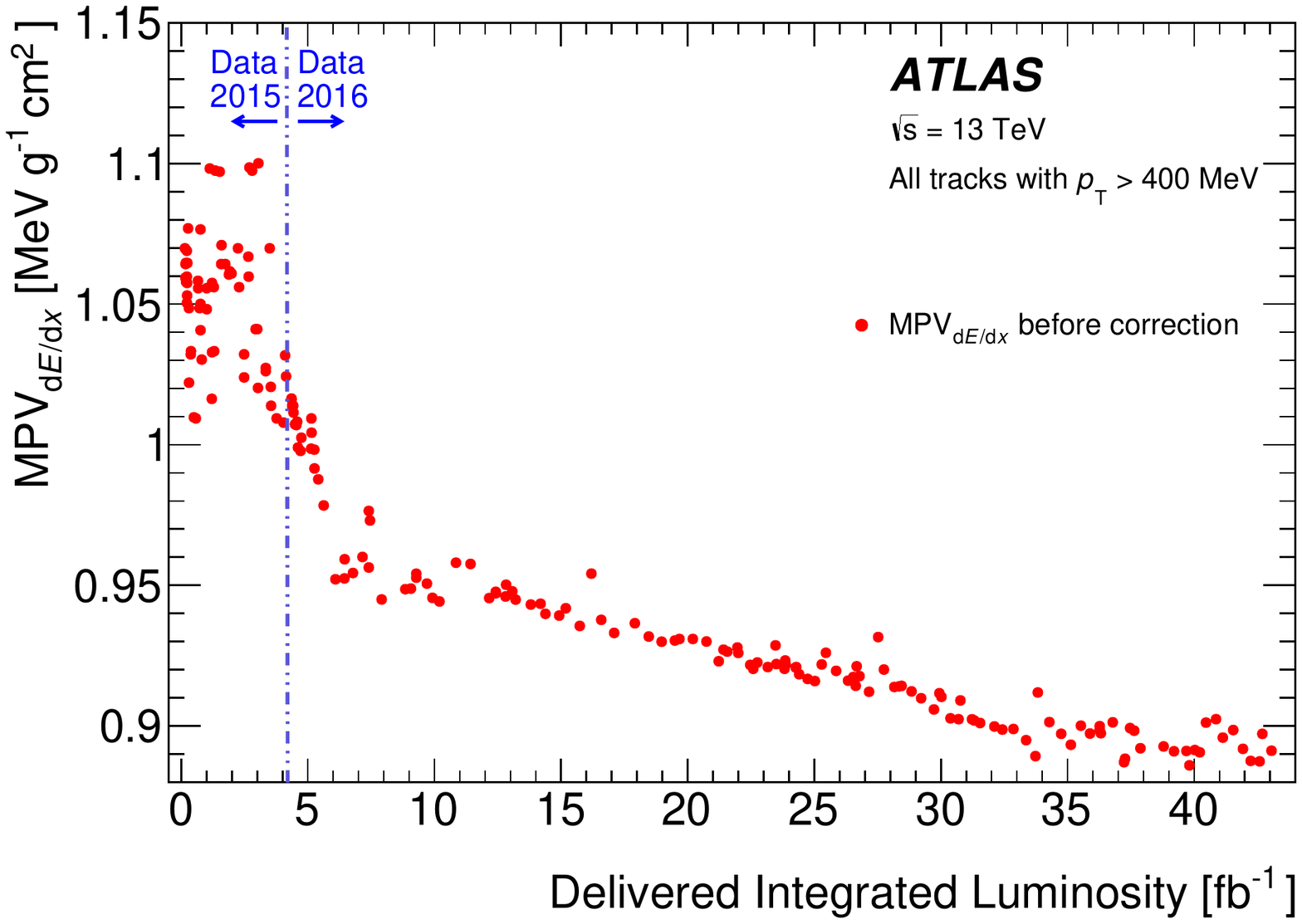}
}
\centering
\caption{The most probable value of the track $\dedx$ (MPV$_{\dedx}$) versus the integrated luminosity delivered to ATLAS is reported for each data-taking run used in this analysis before corrections are applied. The luminosity plotted here is before detector efficiency and data quality criteria are imposed. The MPV$_{\dedx}$ is calculated for all tracks with $p_\text{T} > 400$~\MeV. The points to the left of the dashed line represent the data recorded during 2015, during which a variation of the MPV$_{\dedx}$ due to the ToT drift of the IBL electronics is pronounced. In data recorded during 2016, a drop of MPV$_{\dedx}$ over integrated luminosity is observed due to charge collection efficiency losses. Small local fluctuations are also visible. These are caused by the change of the experimental environment and of the detector conditions. In this analysis, the measurement of \dedx\ is corrected to account for the variation as a function of data-taking run.}

\label{fig:dedx_v_lumi}
\end{figure}

ATLAS has used the measured $\dedx$ to search for $R$-hadrons in several previous analyses~\cite{SUSY-2012-01,SUSY-2014-09,SUSY-2016-03}. This method has been constantly improved to take into account the evolution of the pixel detector and the experimental conditions. Detailed improvements related to the measurement of \dedx\ and mass introduced in this analysis, include:

\begin{itemize}
\item{Corrections have been made for luminosity- and time-dependent variations of the measured values of $\dedx$. The variations are due to changes in the operation parameters of the pixel system and to loss of charge collection due to radiation damage caused by the luminosity delivered. The \dedx\ measured in data is scaled by a per-run factor derived to keep the most probable value of the energy loss (MPV$_{\dedx}$) constant versus time.
The MPV$_{\dedx}$ variation with integrated luminosity before corrections is shown in Figure~\ref{fig:dedx_v_lumi}. }

\item{A low-momentum correction for kaons and protons has been added. All particles are treated as pions in the reconstruction program, but, below 500~\MeV, the effect of multiple scattering on the trajectories of kaons and protons is different from the effect on a pion and their momenta are underestimated. 
To correct for this effect, the difference between the generated and the reconstructed momentum of proton and kaon tracks 
in simulation samples is fitted as a function of momentum. 
This parameterised correction is then applied to the momentum of protons and kaons in data, where these particles are
identified by means of their $\dedx$ and momentum. This procedure has simplified the $\dedx$ calibration, which is performed with low-$\beta\gamma$ SM particles.} 

\item{There is a small dependence of the $\dedx$ on the traversed thickness~\cite{Bichsel88}. For that reason the $\dedx$ calculated in this analysis takes into account its small ($<$10$\%$) $\eta$-dependence. After this correction, the $\dedx$ depends only on the particle momentum and mass, which simplifies the background estimation (see Section~\ref{sec:background}).}

\item{As the simulation does not include the effects of radiation damage to the pixel detector sensors, a scale factor of 0.886 is applied to the measurement of \dedx\ in simulation to align the MPV$_{\dedx}$  of the minimum-ionising particles in MC simulation with data, after the run-dependent corrections to the data \dedx\ have been applied.}
\end{itemize}

\begin{figure}[]
\resizebox{0.65\textwidth}{!}{%
\includegraphics{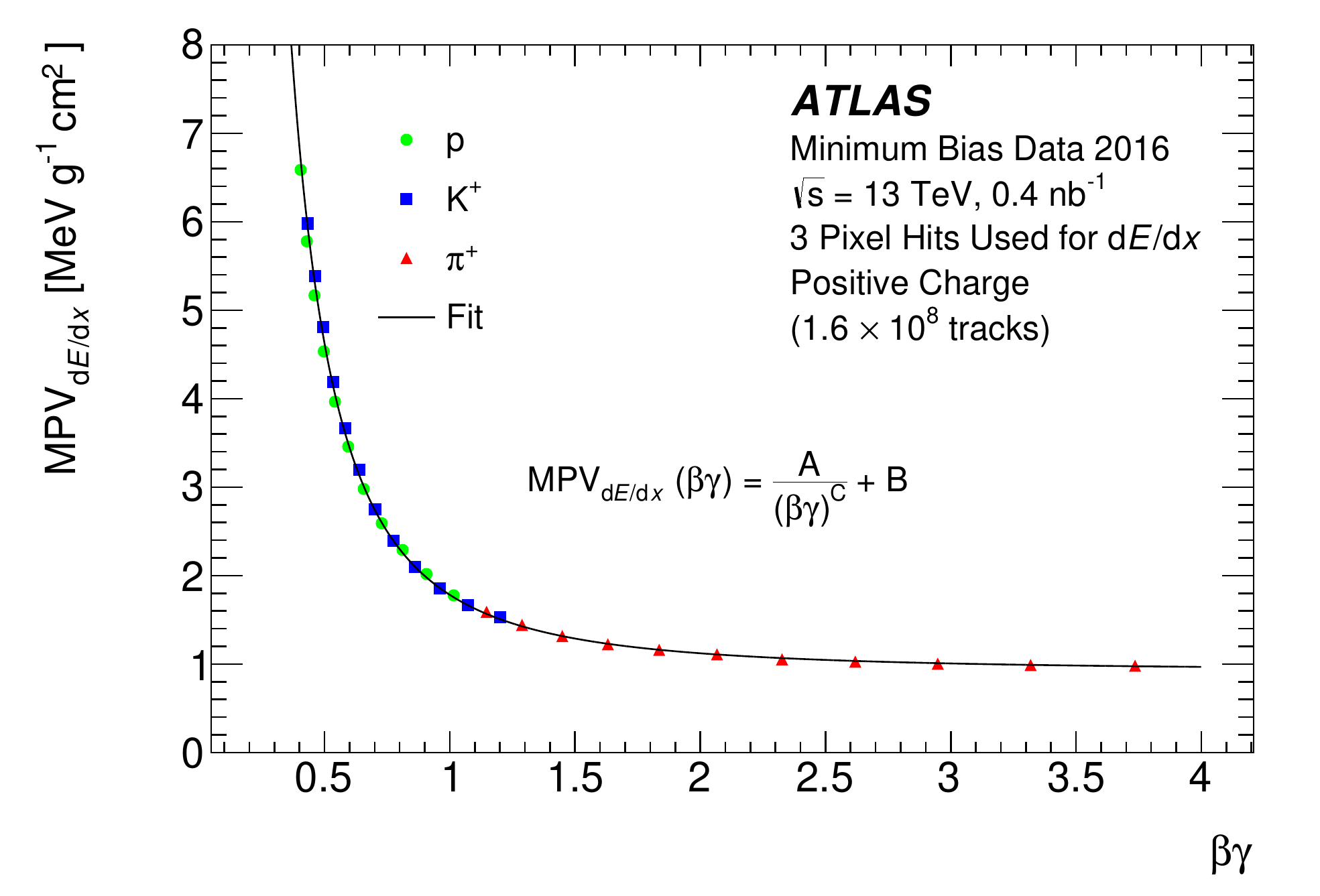}
}
\centering
\caption{MPV$_{\dedx}$ as a function of $\beta\gamma$ obtained with a sample of minimum-bias data from 2016, for positively charged tracks with three pixel hits used to calculate the $\dedx$. This data sample amounts to about 0.4~nb$^{-1}$. For each kaon and proton, the $\beta\gamma$ value is corrected for the effect of multiple scattering. A fit to the MPV$_{\dedx}$ dependence on $\beta\gamma$ with an empirical three-parameter function 
$\textrm{MPV}_{\dedx}= A/{(\beta\gamma)^C+B}$ motivated by Bethe--Bloch relation is also shown. The values of the A, B and C parameters for tracks with different charge and different number of pixel hits are all compatible.}
\label{fig:mpv_v_betagamma}
\end{figure}

The $\beta\gamma$ of a particle, and therefore its mass if the momentum is known, can be calculated from the $\dedx$ of its track using the relationship between $\beta\gamma$ and $\dedx$. A $\beta\gamma$ value can only be measured in the range 0.3~$<\beta\gamma<$~0.9. On average, particles with $\beta\gamma<$~0.3 have a $\dedx$ such that the ToT dynamic range is exceeded. Particles with $\beta\gamma>$~0.9 have a $\dedx$ which is too close to the ionisation plateau of relativistic SM particles for an efficient discrimination. This range overlaps well with the expected average $\beta\gamma$ of $R$-hadrons produced at the LHC, which decreases from around 0.8 for a gluino with mass 600~\GeV\ to around 0.4 for a 2000~\GeV gluino.

The mass of a charged particle can be derived from a fit of the specific energy loss and the momentum measurement to an empirical function motivated by the low-$\beta$ behaviour of the Bethe--Bloch distribution. After applying the low-momentum correction for kaons and protons, it is possible to fit the function relating \dedx\ to $\beta\gamma$ with only three parameters (instead of five as in the previous analysis~\cite{SUSY-2016-03}), as shown in Figure~\ref{fig:mpv_v_betagamma}. The parametric function describing the relationship between the most probable value of the energy loss (MPV$_{\dedx}$) and $\beta\gamma$ is:
\begin{equation}
\textrm{MPV}_{\dedx}= A/{(\beta\gamma)^C+B}
\label{bbfun}
\end{equation}
The $A$, $B$ and $C$ calibration constants were measured using low-momentum pions, kaons and protons reconstructed by ATLAS in low-luminosity runs where all reconstructed tracks with $\pt>$100~\MeV\ are considered. In bins of momentum, the reconstructed $\dEdx$ distribution shows three distinct peaks, to which the nominal pion, kaon, and proton masses are respectively assigned in increasing order of $\dEdx$ to obtain a $\beta\gamma$ for the three measured $\dEdx$ values. The MPV$_{\dedx}$ is extracted from a fit to the distribution of \dedx\ values for each particle species across all momentum bins. The mass parameterisation is valid for both data and simulation after the correction to \dedx\ in simulation is applied.

Given a measured value of $\dEdx$ and momentum, and assuming unit charge, the mass $m$ is calculated from Eq.~(\ref{bbfun}) by numerically solving the equation MPV$_{\dedx}$$(p/m)$ = $\dedx$ for the unknown $m$, where the MPV$_{\dedx}$ is approximated by the truncated-mean measurement of \dedx. Using this method, the reconstructed mass for simulated $R$-hadrons reproduces well the generated mass up to about 1.5~\TeV, above which a bias in the measured momentum causes the reconstructed mass to fall below the generated value. The momentum uncertainty dominates the mass resolution above masses of 200~\GeV. The measurement of the proton mass in all data-taking runs used in this analysis allows the monitoring of the stability of the $A$, $B$ and $C$ calibration constants. These are found to be stable at the 1\% level after all corrections have been applied.

%-------------------------------------------------------------------------------
\section{Event selection}
\label{sec:selection}
%-------------------------------------------------------------------------------

Events are first selected with a trigger based on \MET, which is calculated using energy measurements in the calorimeter with corrections for multiple $pp$ interactions in each event~\cite{TRIG-2016-01}. The high-level \MET trigger threshold varies from 70~\GeV\ to 110~\GeV\ during the data-taking period. In the reconstruction, \MET is built from calibrated muons and electrons which pass baseline selections, from calibrated jets reconstructed with the anti-$k_t$ jet clustering algorithm~\cite{antikt} with radius parameter $R=0.4$ using clusters of energy depositions in the calorimeter as inputs, and from a term that includes soft tracks not associated with any other objects in the event~\cite{PERF-2014-04} but consistent with the primary vertex (PV). Events are required to have  $\MET~>~170$~\GeV\ to enhance the signal sensitivity and to ensure that the selected events are near the efficiency plateau of the trigger. To ensure a good calculation of \MET, events are rejected if they contain a jet with $E_{\text{T}} > 20$~\GeV\ that is consistent with detector noise or beam-induced backgrounds, as determined from shower shape information. Unlike in standard ATLAS selections for jet-cleaning~\cite{ATLAS-CONF-2015-029}, a requirement on the relationship between track and calorimeter measurements of $p_{\text{T}}$ and a requirement on the fraction of jet energy deposited in the electromagnetic calorimeter are not applied as they are found to be inefficient for signal events in which an $R$-hadron decays before or inside the calorimeters. 
The trigger is more than 95\% efficient for $R$-hadrons with lifetimes of 10~ns or less; the efficiency decreases as more decays happen in or after the calorimeter and falls to around 30-40\% for the stable case.

%For signal events which pass the offline \MET requirement, the trigger is more than 95\% efficient for $R$-hadrons with a lifetime of 10~ns and around 70\% efficient for the stable case.

There are two separate signal regions with slightly different optimisations for metastable and stable particles: the isolation selections differ slightly for the two signal regions, and a muon veto is applied only for the metastable region. Additionally, events with a high-$p_{\text{T}}$ muon whose momentum uncertainty is significantly worse after combining tracks from the inner detector and muon system are vetoed in the metastable region, in order to protect the measurement of \MET from rare, pathological reconstructions of muons.  After passing the trigger and \MET selections, events are required to have a PV built from at least two reconstructed tracks each with $p_{\text{T}}$ above 400~\MeV, and to contain at least one candidate track that passes the track-level selections detailed below. If there are multiple candidate tracks in an event after all selections, the candidate with the highest track $p_{\text{T}}$ is selected.

To enrich the selected sample in potential signal events, candidate tracks are required to have $p_\text{T} > 50$~\GeV, momentum $p > 150$~\GeV, and $|\eta| < 2.0$. To reject non-prompt background tracks and those inconsistent with the PV, the transverse impact parameter of candidate tracks, $|d_0|$, must be less than 2~mm, and the absolute value of the product of the longitudinal impact parameter, $z_0$, and $\sin\theta$ must be less than 3~mm.\footnote{The transverse impact parameter is defined as the distance of closest approach in the transverse plane between a track and the beam-line. The longitudinal impact parameter corresponds to the $z$-coordinate distance between the point along the track at which the transverse impact parameter is defined and the primary vertex.} Reconstructed tracks must have at least seven clusters across the pixel and SCT detectors, and to be considered a candidate the track must have an associated cluster in the innermost pixel layer if it passes through an active detector module. 

To reject tracks from leptonic $W$ decays, the transverse mass ($m_\text{T}$) of the candidate track and the \MET in the event must be greater than 130~\GeV.\footnote{$m_\text{T} = \sqrt{2 \pt \MET \left( 1-\cos(\Delta\phi (\MET,\text{track})) \right)}$, where $\Delta\phi (\MET,\text{track})$ is the azimuthal separation between the track and the \MET\ vector.} Tracks from electrons are removed by considering any jets within $\Delta R = 0.05$ of the candidate track with $p_\text{T} > 20$~\GeV, and rejecting the track if any such jet has at least $95\%$ of its energy deposited in the electromagnetic calorimeter. SM hadrons are removed by excluding tracks for which any associated jet within $\Delta R = 0.05$ of the track has a calibrated energy larger than the track momentum. In the metastable $R$-hadron signal region, tracks identified as well-reconstructed muons which pass the ``medium'' quality selection~\cite{PERF-2015-10} and which have $p_\text{T} > 25$~\GeV\ are rejected.

Tracks with high ionisation deposits from multiple SM particles which overlap in the pixel sensors are rejected with two types of isolation selections. The first explicitly requires that no clusters on the track are consistent with two or more tracks~\cite{PERF-2015-08}. The second requires that the scalar sum of the $p_\text{T}$ of other tracks, with $p_\text{T} > 1$~\GeV\ and consistent with the PV, in a cone of size $\Delta R = 0.25$ around the candidate track, must be less than 20~\GeV\ for the metastable $R$-hadron selection. To reduce background in the stable $R$-hadron region in which muons are not vetoed, the isolation selection is tightened to 5~\GeV.

At least two pixel clusters, after discarding the cluster with the highest ionisation, must be included in the truncated mean calculation of \dedx\ to ensure it is robust. The relative uncertainty in the momentum measurement must be less than $50\%$. The specific ionisation of the candidate track measured by the pixel detector must be larger than $1.8$~\MeV~g$^{-1}$~cm$^{2}$. Relative to inclusive generated $R$-hadron events with a mass of 2000~\GeV, the efficiency for events to pass all selections, including the trigger, is 12\% for stable $R$-hadrons and 19\% for those with a lifetime of 10~ns.

%make sure detector section defines:
%hits, clusters
%corrections to \dedx measurement

%make sure analysis overview defines both selection regions

%-------------------------------------------------------------------------------
\section{Background estimation}
\label{sec:background}
%-------------------------------------------------------------------------------

The expected background contains tracks from SM processes including vector boson, top-quark, and multi-jet production. Tracks from any SM particle can be measured with high \dedx\ due to the unlikely sampling of multiple measurements from the long tail of the Landau distribution, from overlapping particles depositing charge in the same pixels, or from spurious pixel hits from low-momentum particles being incorrectly assigned to the high-momentum track. To correctly estimate both the rate of high-momentum tracks in events with large \MET and the probability of measuring a high ionisation energy for those tracks, the background is fully estimated from data.

%Tracks from any high momentum SM particle can be measured with high \dedx\ due to various low probability occurrences.
%For instance the multiple ionisation measurements may all be from the long tail of the Landau distribution, or overlapping particles deposit charge in the same pixel cluster or spurious clusters from a low momentum track are associated to a high momentum track. As all the above phenomena can hardly be simulated in a satisfactory way, the background is fully estimated from data.

A template for the momentum distribution of background tracks in signal region (SR) events is obtained from a control region ($p$-CR) in which the
ionisation requirement is inverted, $\dedx < 1.8$~\MeV~g$^{-1}$~cm$^{2}$, while all other track-level and event-level selections
are applied. 

The \dedx\ distribution, in a few bins of momentum\footnote{to account for the
dependence of \dedx\ on momentum up to the Fermi plateau. The most probable energy loss reaches a constant value, the Fermi plateau, at large $\beta\gamma$.}, is obtained for the expected background from a low-\MET data sample in which $\MET~<~170$~\GeV. Inverting the \MET requirement relative to the high-\MET SR
minimises signal contamination in this control region (\dedx-CR), and the lack of
correlation between \MET and \dedx\ for high-momentum SM tracks allows
the \dedx\ distribution of the expected background to be derived from
low-\MET events which pass all other selections. Since the \MET trigger
thresholds varied as a function of time for the collected data, the
events in this control region are reweighted so that the ratio of
low-to-high \MET events is constant versus time.

The momentum and \dedx\ distributions obtained in the control regions (CR)s are used as templates to calculate the shape of the expected mass distribution of candidate tracks from background events. A pair of $p$ and \dedx\ values is obtained by randomly sampling from the $p$-CR
distribution, and then randomly sampling from the \dedx-CR distribution 
in the appropriate $p$-bin. The mass for each pair of $p$
and \dedx\ values is calculated as described in
Section~\ref{sec:mass}. The resulting background mass distribution
is normalised to data in the region where $m < 160$~\GeV, in which
signal was previously excluded~\cite{SUSY-2012-01,Acosta:2002ju}, before the high ionisation
requirement is imposed.

\begin{figure}[htpb]
\centering
\subfloat[]{\includegraphics[width=0.45\textwidth]{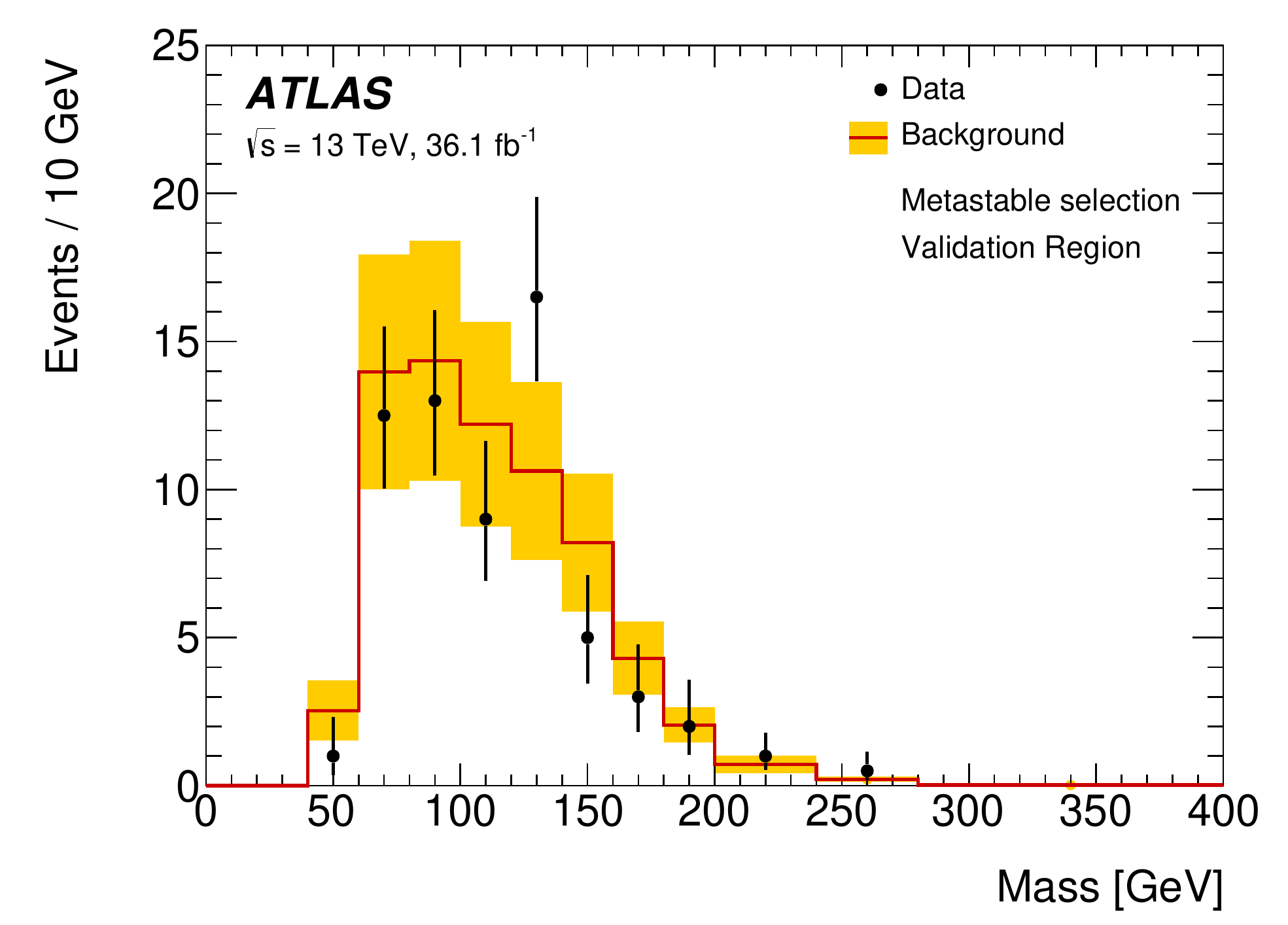}}
\subfloat[]{\includegraphics[width=0.45\textwidth]{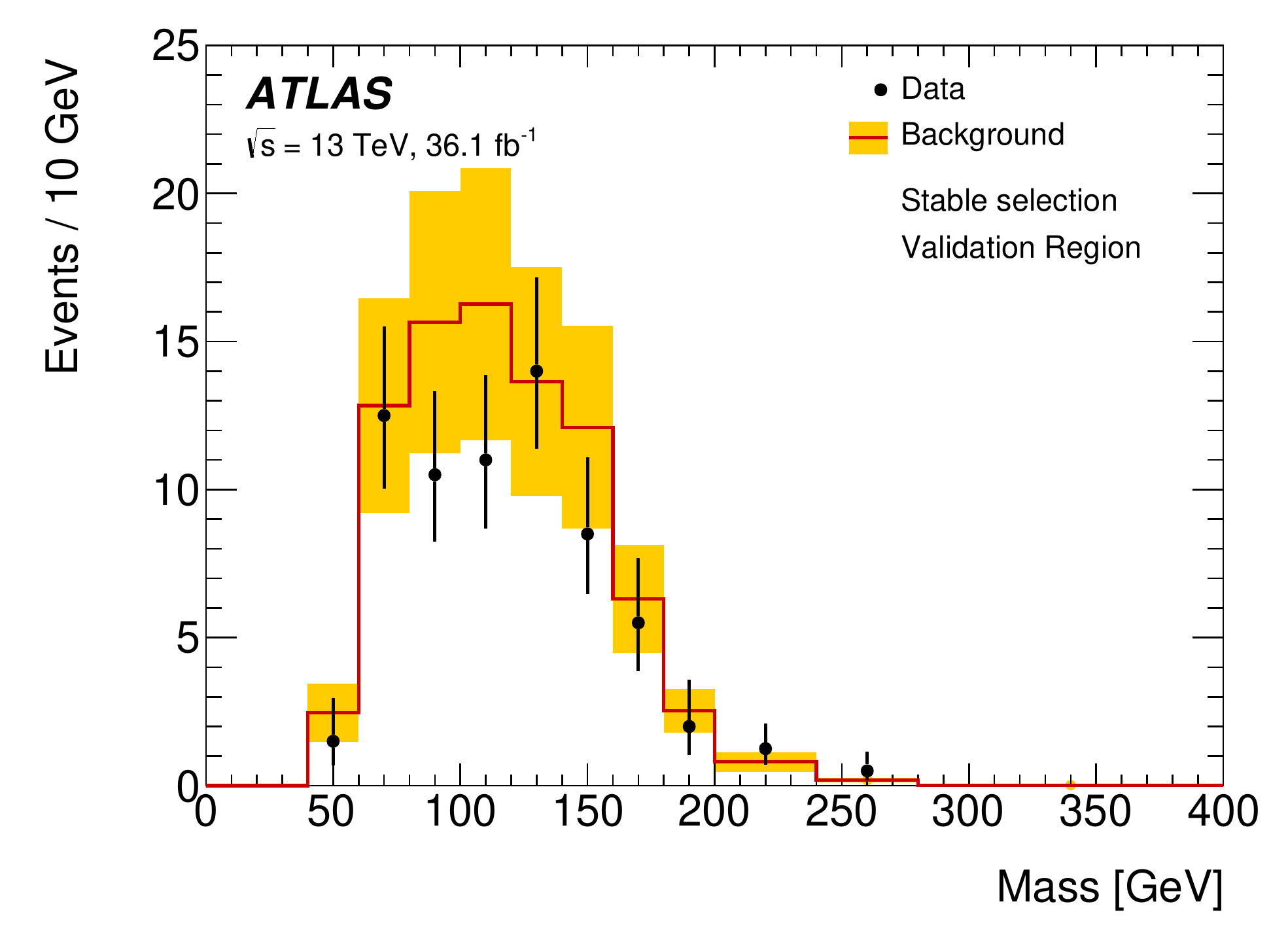}}
\caption{The reconstructed mass distribution in the (a) metastable and (b) stable $R$-hadron
  validation regions for observed data and the predicted
  background, including the total uncertainty in the background estimate. The validation regions have
  the same requirements as the SRs, except the momentum of the candidate
  tracks is required to be $50 < p< 150$~\GeV.}
\label{fig:background_validation}
\end{figure}

The procedure for estimating both the normalisation and shape of the
expected background is validated in a low-momentum validation region (VR) in which the
momentum of tracks is required to be between 50~\GeV\ and 150~\GeV. The differences between the selections applied to the SR, CR, and VR are shown in Table~\ref{table:regions}. The control and validation regions are independently produced for both the metastable and stable $R$-hadron SRs. The expected mass distributions in the two validation regions, along with the observed data, are shown in Figure~\ref{fig:background_validation}. Good agreement between the data and the prediction in the VR validates the background estimation procedure.

\begin{table}[htbp]
\centering
\caption{Summary of the different selection requirements applied to the signal region (SR), the validation region (VR), and the control regions (CR).}
\begin{tabular}{|r|c|c|c|c|c|c|}
\hline

                                                     & \multirow{2}{*}{SR} & \multirow{2}{*}{VR}    &  \multicolumn{2}{c|}{$p$-CR}         & \multicolumn{2}{c|}{\dedx-CR}\\
                                       & &  & for SR     & for VR           & for SR     & for VR\\
\hline
Track Momentum [\GeV]                           & $>$150        &  {50--150}      & $>$150        &  {50--150}               & $>$150        &  {50--150}       \\
\met\ [\GeV]                           & \multicolumn{2}{c|}{ $>$170} &\multicolumn{2}{c|}{ $>$170}       &  \multicolumn{2}{c|}{ {$<$170} }           \\ 
Ionisation [\MeV\ g$^{-1}$ cm$^2$]  &  \multicolumn{2}{c|}{$>$ 1.8}         &   \multicolumn{2}{c|}{{$<$ 1.8}}         &   \multicolumn{2}{c|}{--}\\
\hline
    \end{tabular}
\label{table:regions}
\end{table}

%-------------------------------------------------------------------------------
\section{Systematic uncertainties}
\label{sec:uncertainties}
%-------------------------------------------------------------------------------

The background estimation technique described in the previous section relies on the lack of correlation between several key kinematic variables in background events. The largest uncertainties in the central value of the background estimate come from possible residual correlations. In particular, the residual correlation between $\eta$ and \dedx\ results in an uncertainty in the size of the background estimate ranging from $15\%$ at the lowest mass values to $30\%$ at the highest mass values. This uncertainty is assessed by comparing the nominal background estimate with an estimate performed in $\eta$ bins. Additionally, an uncertainty of $1\%$--$25\%$ in the background yield arises from residual correlations between $p$ and \dedx\ for tracks entering the background calculation. This is estimated by reweighting the $p$ template from the $p$-CR by the difference in the $p$ distribution between tracks with high and low \dEdx\ in the low-\MET region. Similarly, the residual correlation between \MET and \dedx\ is probed by rescaling the template \dEdx\ distribution with a scale factor obtained from the difference between the \dedx\ distributions in the VR for tracks in events with high \MET and low \MET. This uncertainty ranges from $3\%$ to $12\%$ on the background expectation in different mass windows.

As the background is fully estimated from data, detector or data-taking conditions which affect the measurement of \dedx\ are accounted for, as long as the luminosity profile of the control regions matches that of the signal region. The reweighting of the \dedx-CR control region achieves this. A conservative uncertainty in the time-dependence of the \dEdx\ measurement is assessed by comparing the background estimate with and without the reweighting, which results in an additional uncertainty of $3\%$--$18\%$ on the background yields. The limited numbers of events in the control regions contribute $6\%$ uncertainty. Other uncertainties in the background estimate are below $5\%$, including an uncertainty in the shape of the \dedx\ tail from the CR and in the different fractions of muons between the CR and SR.

The uncertainty in the expected number of signal events is dominated by the estimation of the production cross-section of gluino--gluino pairs; the calculation of the cross-section and its uncertainty is described in Section~\ref{sec:samples}. The uncertainty ranges from $14\%$ for gluino masses of 600~\GeV\ to $36\%$ for masses of 2200~\GeV. An additional uncertainty in the number of produced signal events of $2.1\%$ is due to the uncertainty in the dataset luminosity, which is measured in dedicated $x$--$y$ beam-separation scans performed in May 2016 using a method similar to one described in Ref.~\cite{DAPR-2013-01}.

The largest uncertainty on the signal efficiency results from the modelling of ISR production, which affects the \MET distribution. This uncertainty is estimated as half the difference between the expected number of events calculated with the \textsc{Pythia 6.4.27}  gluino--gluino $p_\textrm{T}$ distribution and with the distribution reweighted to match that of the \textsc{MadGraph5}{\_a}\textsc{MC@NLO} sample. This uncertainty depends on both the lifetime and mass of the signal sample and ranges from $1\%$ for lifetimes up to 10~ns to $19\%$ for stable samples. Uncertainties ranging from $1\%$ to $6\%$ in the efficiency of the \dedx\ selection are included to account for both the shape difference between the ionisation distributions in data and MC simulation and the scale shift in data due to radiation damage. The efficiency of selecting tracks with at least two measurements used to determine the \dedx\ depends on the operating conditions of the detector and the instantaneous luminosity. The accuracy of the simulation in modelling this efficiency is tested in $Z\rightarrow\mu\mu$ events in both data and MC simulation; the maximum difference in efficiency as a function of pile-up is found to be $6\%$, which is taken as an uncertainty. Additional uncertainties, each less than $5\%$, on the signal selection efficiency are due to uncertainties in how well the simulation models trigger and offline \MET, the pile-up distribution, the scale and uncertainty of the momentum measurement, and the efficiency for reconstructing stable $R$-hadrons as muons.

%-------------------------------------------------------------------------------
\section{Results}
\label{sec:results}
%-------------------------------------------------------------------------------

\begin{figure}[htbp]
  \centering 
  \subfloat[]{\includegraphics[width=0.65\textwidth]{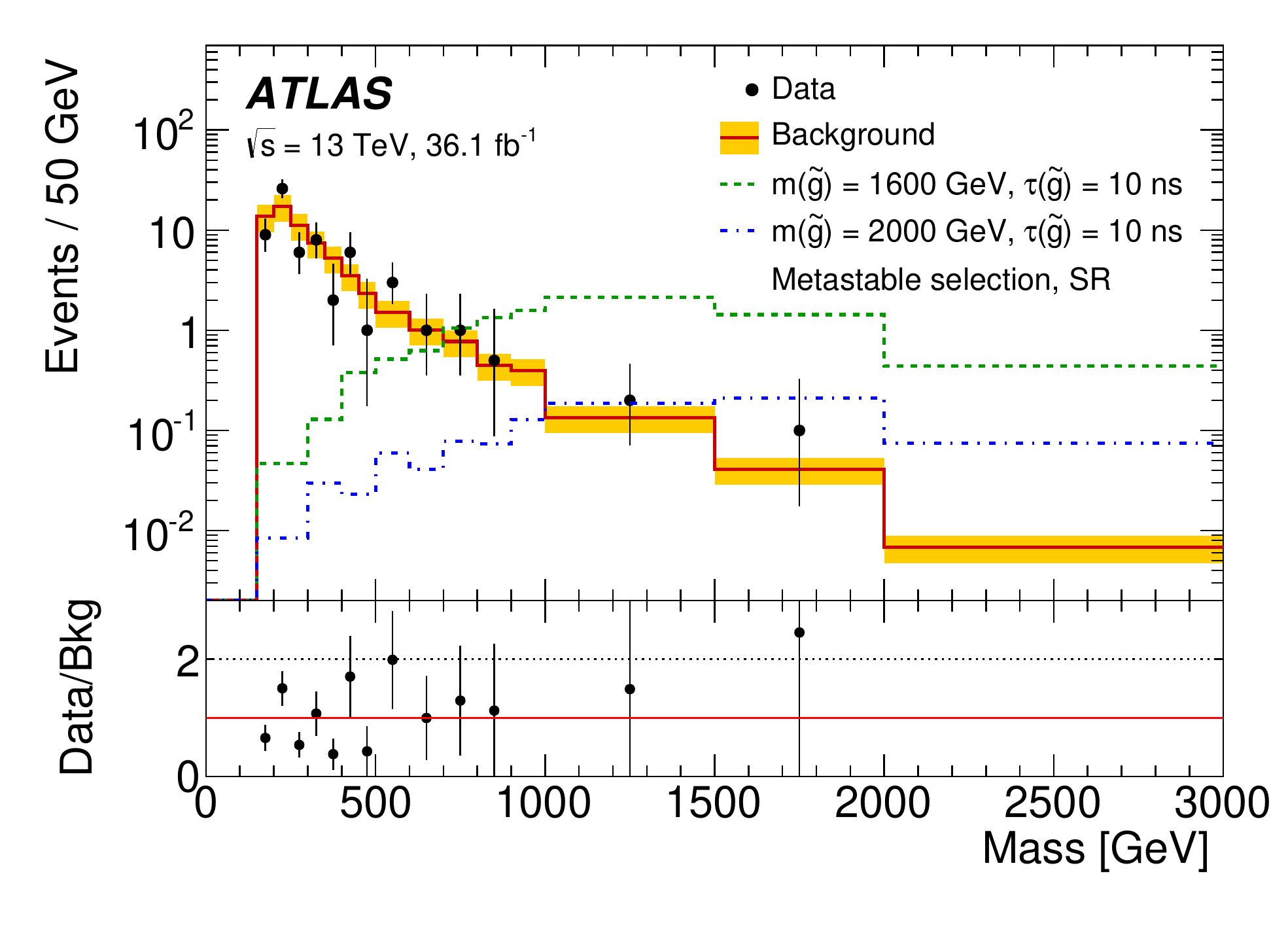}}

  \subfloat[] {\includegraphics[width=0.65\textwidth]{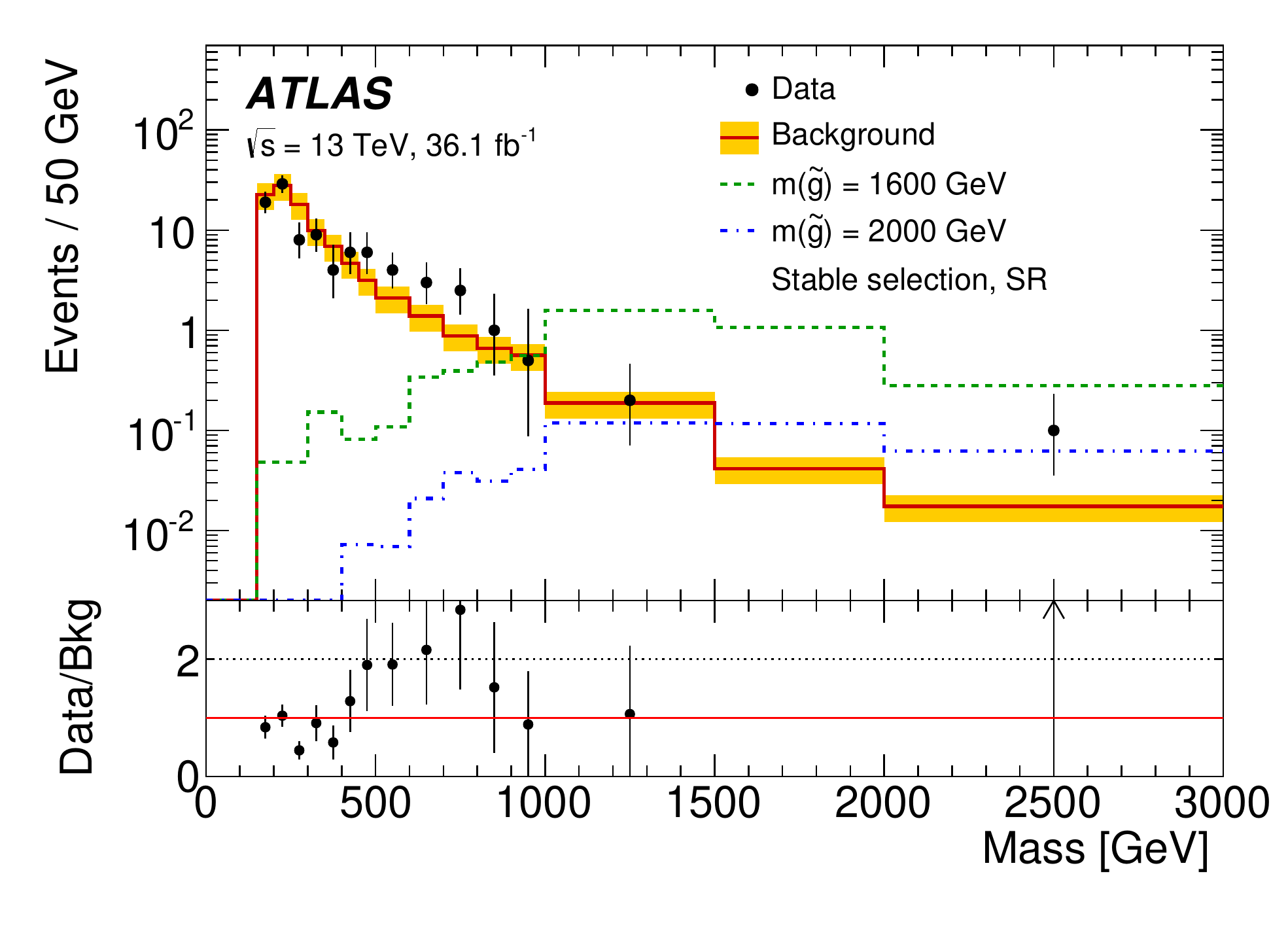}}
  \caption{The reconstructed candidate track mass distributions for observed data,
    predicted background, and the expected contribution from two signal models in
    the (a) metastable and (b) stable $R$-hadron signal regions. The yellow band around the background estimation includes both the statistical and systematic uncertainties.}
  \label{fig:results}
\end{figure}

The distributions of the reconstructed mass of candidate tracks in
the two signal regions are shown in Figure~\ref{fig:results} for events observed in data, together with the
expected background and the predictions from several signal models. The total numbers of
expected and observed events in the two SRs as well as in the background CRs
and VRs are shown in Table~\ref{table:results}. Overall, the number of
observed events in the two SRs is consistent with the background expectation.

\begin{table}[htbp]
\centering
    \caption[]{The number of events in each CR, VR,
      and SR for the predicted background, for the expected
      contribution from two signal models
      normalised to $\lumi$~fb$^{-1}$, and in the observed data. The
      predicted background includes the statistical and systematic
      uncertainties, respectively. The uncertainty in the
      signal yield includes all systematic uncertainties except that in the theoretical cross-section.}
    \begin{tabular}{|c|l|c|c|r|}
       \hline
Region & Sample & Pred. Bkg ($\pm$ stat. $\pm$ syst.) & Exp. Signal & Data \\
\hline
\multirow{5}{*}{Metastable} & & & \underline{$m({\tilde{g})}=1600~\GeV$, $\tau({\tilde{g})} = 10$~ns} & \\
%\cline{4}
& $p$-CR & - & 12.0 $\pm$ 0.9 & 7397 \\
& \dedx-CR & - & 7.2 $\pm$ 0.6 & 110019 \\
& VR & 140 $\pm$ 4 $\pm$ 28 & 0.3 $\pm$ 0.03 & 130 \\  
\hdashline
& SR & 71 $\pm$ 2 $\pm$ 14& 52.1 $\pm$ 4.2 & 72 \\
\hline
\multirow{5}{*}{Stable} & & & \underline{$m({\tilde{g})}=1600~\GeV$, stable} & \\
%\cline{4}
& $p$-CR & - & 8.0 $\pm$ 1.6 & 13108 \\
& \dedx-CR &- & 10.3 $\pm$ 2.1 & 272723 \\
& VR & 168 $\pm$ 5 $\pm$ 32 & 0.2 $\pm$ 0.04 & 138 \\ 
\hdashline
& SR & 107 $\pm$ 3 $\pm$ 28 & 36.0 $\pm$ 7.2 & 107 \\
\hline
    \end{tabular}
\label{table:results}
\end{table} 

To quantify the level of agreement between data and background in the shape of the mass distribution, discrete but overlapping asymmetric windows in the reconstructed mass distribution are defined so as to contain at least 70\% of the reconstructed mass of a
signal sample with a given simulated gluino mass. All windows have an upper
boundary of 5000~\GeV\ to remove any unphysical measurements. The lower boundary for a given simulated mass
varies slightly for different lifetimes. Twelve windows are used in each of the two signal regions. The compatibility of the
observed event counts with the background expectation is tested within
each mass window. The largest deviation from the background-only
hypothesis is found to have a local significance of $2.4\sigma$ and is in the stable $R$-hadron SR in the mass bin 
designed to cover a 600~\GeV\ gluino (with a mass window from 500 to 5000~\GeV). The source of this deviation is a mild excess of data events relative to the background prediction around 500--800~\GeV.

In the absence of any significant excess, model-independent upper
limits at 95\% CL on the visible production cross-sections are
calculated by dividing the number of signal events consistent at the
95\% CL with the expected background and observed data in the most
inclusive mass window for each SR by the integrated luminosity. For
the metastable $R$-hadron SR, the $p$-value for the background-only
hypothesis is 0.15 in the window from 500 to 5000~\GeV, and the upper
limit on the visible production cross-section is $0.35$~fb with an
expected limit of $0.25^{+0.09}_{-0.07}$~fb. In the stable $R$-hadron
SR mass window from 300 to 5000~\GeV, the background-only $p$-value is
0.09 and the model-independent upper limit on the visible production
cross-section is $0.88$~fb, with an expected limit of
$0.57^{+0.20}_{-0.12}$~fb.  Information in full detail about the expected and observed results in each mass window is provided in Ref.~\cite{Hepdata}.

\begin{figure}[htbp]
  \centering 
  \subfloat[] {\includegraphics[width=0.45\textwidth]{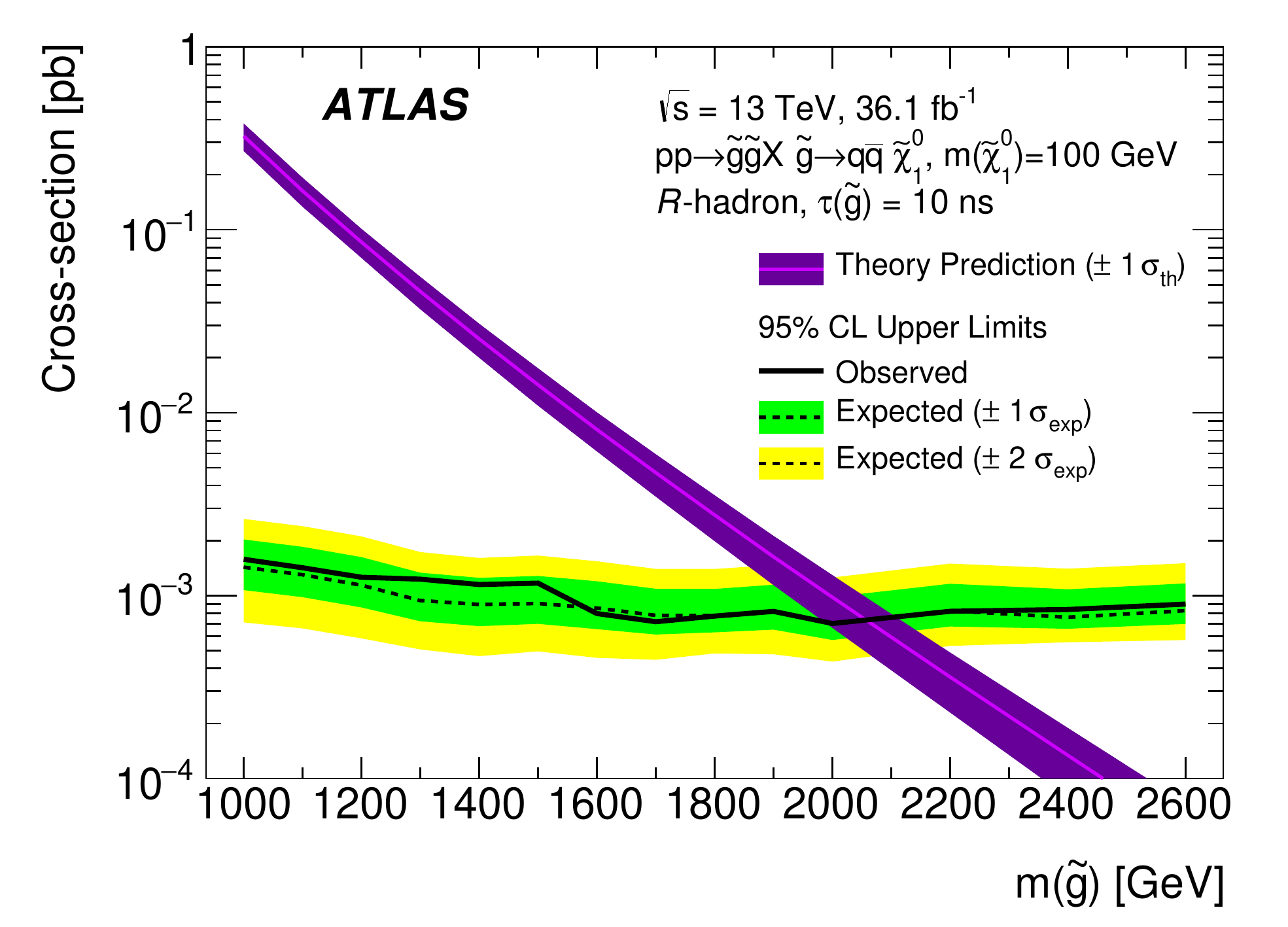}}
  \subfloat[] {\includegraphics[width=0.45\textwidth]{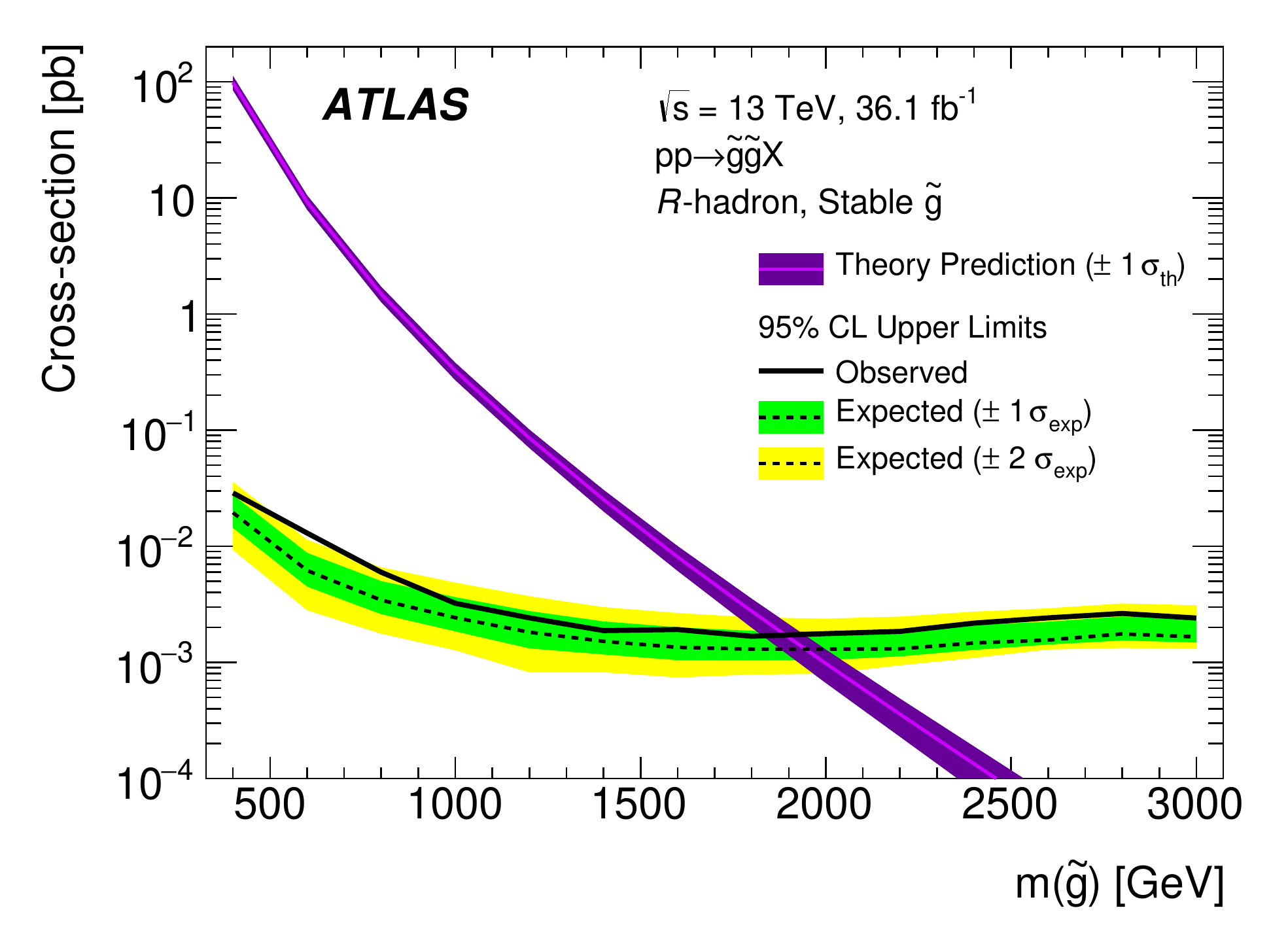}}
  \caption{The 95\% CL upper limit on the cross-section as a function of
mass for (a) gluinos with lifetime $\tau = 10$~ns decaying into $q\bar{q}$ and a
100~\GeV\ neutralino and for (b) detector-stable gluinos, with the observed limit shown as a solid black line. The predicted production cross-section values are
shown in purple along with their uncertainty. The expected upper limit in the
case of only background is shown by the dashed black line, with a green
$\pm 1\sigma$ and a yellow $\pm 2\sigma$ band. Theory cross-sections are from Refs.~\cite{Beenakker:1996ch,Kulesza:2008jb,Kulesza:2009kq,Beenakker:2009ha,Beenakker:2011fu}.}
  \label{fig:crosssection_limit}
\end{figure}

Expected and observed upper limits on $R$-hadron production
cross-sections are calculated from the predicted background, the expected signal, and the observed event
yields in each mass window, using the one-sided profile-likelihood ratio as a
test statistic. The upper limits on the cross-sections are evaluated
at 95\% CL following the CL$_\text{s}$ prescription~\cite{Read:2002hq}. In this procedure, the uncertainties in the
signal and background yields are treated as Gaussian-distributed
nuisance parameters. The cross-section upper limits for a gluino $R$-hadron with lifetime of 10~ns decaying into $q\bar{q}$ and a 100~\GeV\
neutralino and for a detector-stable $R$-hadron are shown in Figure~\ref{fig:crosssection_limit}. 

\begin{figure}[htbp]
  \centering 
  \resizebox{0.65\textwidth}{!}{%  
    \includegraphics{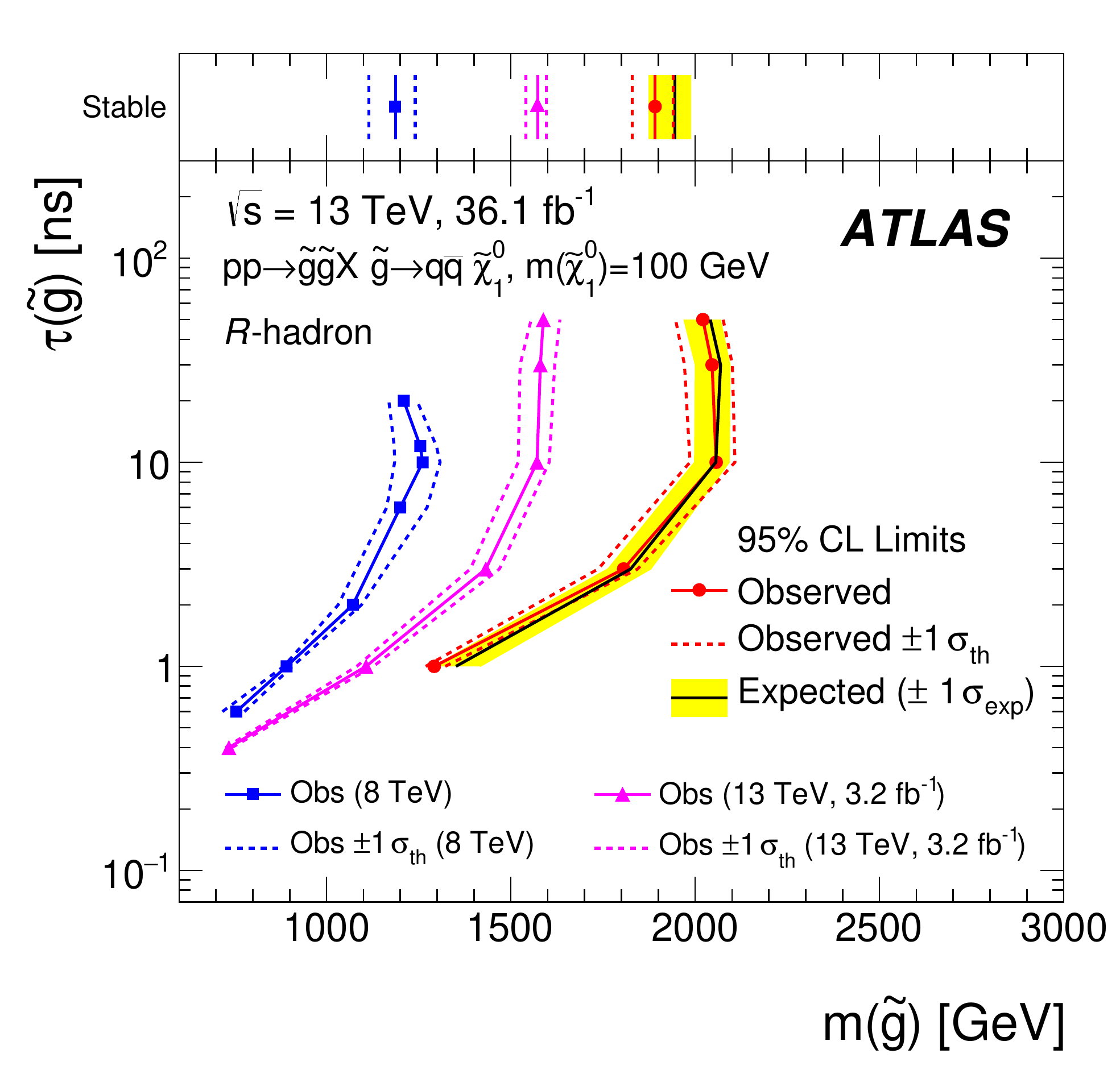}
  }
  \caption{Observed and expected $95\%$ lower limits on the gluino mass in the gluino lifetime--mass plane. The excluded area is to the left of the curves. The observed limit is shown by the solid red line with dot markers with $\pm 1\sigma$ of its theoretical uncertainties ($\sigma_{\text{th}}$) shown as dashed-red lines, and the expected limit
is shown as a black line with $\pm 1\sigma$ of its experimental uncertainties ($\sigma_{\text{exp}}$) shown as a yellow band. The 8~\TeV\ results, shown with blue squares, are from Ref.~\cite{SUSY-2014-09} and the 13~\TeV\ results with 3.2~fb$^{-1}$, shown with pink triangles, are from Ref.~\cite{SUSY-2016-03}.}
\label{fig:lifetime_v_mass}
\end{figure}

The cross-section limits and the predicted production cross-sections for gluinos are used to set lower limits on
expected and observed masses, as a function of
lifetime. The excluded regions in the lifetime--mass plane for gluino
$R$-hadrons which decay into a 100~\GeV\ neutralino and quarks are
shown in Figure~\ref{fig:lifetime_v_mass}. Masses smaller than
2060~\GeV\ are excluded for the most sensitive lifetime of 10~ns, masses smaller than 1890~\GeV\ are excluded for the stable case, and masses smaller than 1290~\GeV\ are excluded for a lifetime of 1~ns. Sensitivity to signals with lifetimes shorter than 1~ns falls off quickly, and is complemented by searches for disappearing tracks~\cite{Aaboud:2017mpt} and displaced vertices~\cite{SUSY-2016-08}. The selection and trigger efficiency, and therefore mass sensitivity, is comparable for a wide range of neutralino masses. For neutralino masses that approach the mass of the gluino, the total efficiency drops by up to a factor of three in these highly compressed decays.

\FloatBarrier

%-------------------------------------------------------------------------------
\section{Conclusion}
\label{sec:conclusion}
%-------------------------------------------------------------------------------

A search has been performed for stable and metastable non-relativistic long-lived particles produced in \emph{pp} collisions at $\sqrt{s}=13$~\TeV\ at the LHC and identified through their large momenta and anomalous specific ionisation energy loss in the ATLAS pixel detector. The data sample analysed corresponds to an integrated luminosity of $\lumi$~fb$^{-1}$ collected by the \mbox{ATLAS} experiment in 2015 and 2016. 
Results are interpreted assuming the pair production of $R$-hadrons as composite colourless states of a long-lived gluino and SM partons.
With some model-dependent assumptions, a lifetime-dependent lower limit is set on the mass of metastable and stable gluinos inside $R$-hadrons. Maximum sensitivity is reached for gluinos with lifetimes of 10~ns, for which masses smaller than 2060~\GeV\ are observed to be excluded at the 95\% confidence level.
%with an expected sensitivity of 2060~GeV. 
Stable gluinos with masses smaller than 1890~\GeV\ are excluded at 95\% confidence level.
% with an expected sensitivity of 1950~GeV.

%-------------------------------------------------------------------------------
\section*{Acknowledgements}
%-------------------------------------------------------------------------------

% Acknowledgements for papers with collision data
% Version 14-Feb-2018

% Standard acknowledgements start here
%----------------------------------------------
We thank CERN for the very successful operation of the LHC, as well as the
support staff from our institutions without whom ATLAS could not be
operated efficiently.

We acknowledge the support of ANPCyT, Argentina; YerPhI, Armenia; ARC, Australia; BMWFW and FWF, Austria; ANAS, Azerbaijan; SSTC, Belarus; CNPq and FAPESP, Brazil; NSERC, NRC and CFI, Canada; CERN; CONICYT, Chile; CAS, MOST and NSFC, China; COLCIENCIAS, Colombia; MSMT CR, MPO CR and VSC CR, Czech Republic; DNRF and DNSRC, Denmark; IN2P3-CNRS, CEA-DRF/IRFU, France; SRNSFG, Georgia; BMBF, HGF, and MPG, Germany; GSRT, Greece; RGC, Hong Kong SAR, China; ISF, I-CORE and Benoziyo Center, Israel; INFN, Italy; MEXT and JSPS, Japan; CNRST, Morocco; NWO, Netherlands; RCN, Norway; MNiSW and NCN, Poland; FCT, Portugal; MNE/IFA, Romania; MES of Russia and NRC KI, Russian Federation; JINR; MESTD, Serbia; MSSR, Slovakia; ARRS and MIZ\v{S}, Slovenia; DST/NRF, South Africa; MINECO, Spain; SRC and Wallenberg Foundation, Sweden; SERI, SNSF and Cantons of Bern and Geneva, Switzerland; MOST, Taiwan; TAEK, Turkey; STFC, United Kingdom; DOE and NSF, United States of America. In addition, individual groups and members have received support from BCKDF, the Canada Council, CANARIE, CRC, Compute Canada, FQRNT, and the Ontario Innovation Trust, Canada; EPLANET, ERC, ERDF, FP7, Horizon 2020 and Marie Sk{\l}odowska-Curie Actions, European Union; Investissements d'Avenir Labex and Idex, ANR, R{\'e}gion Auvergne and Fondation Partager le Savoir, France; DFG and AvH Foundation, Germany; Herakleitos, Thales and Aristeia programmes co-financed by EU-ESF and the Greek NSRF; BSF, GIF and Minerva, Israel; BRF, Norway; CERCA Programme Generalitat de Catalunya, Generalitat Valenciana, Spain; the Royal Society and Leverhulme Trust, United Kingdom.

The crucial computing support from all WLCG partners is acknowledged gratefully, in particular from CERN, the ATLAS Tier-1 facilities at TRIUMF (Canada), NDGF (Denmark, Norway, Sweden), CC-IN2P3 (France), KIT/GridKA (Germany), INFN-CNAF (Italy), NL-T1 (Netherlands), PIC (Spain), ASGC (Taiwan), RAL (UK) and BNL (USA), the Tier-2 facilities worldwide and large non-WLCG resource providers. Major contributors of computing resources are listed in Ref.~\cite{ATL-GEN-PUB-2016-002}.
%----------------------------------------------

%-------------------------------------------------------------------------------
\clearpage

%-------------------------------------------------------------------------------
% If you use biblatex and either biber or bibtex to process the bibliography
% just say \printbibliography here
\printbibliography
%-------------------------------------------------------------------------------

\clearpage % ATLAS Collaboration author list
% Reference date of SUSY-2016-31 is 2018-05-23
% Author list last updated on date 12-AUG-18
% Data extracted on 12-Aug-2018 for paper reference SUSY-2016-31
% at 12:18pm
 
\begin{flushleft}
{\Large The ATLAS Collaboration}

\bigskip

M.~Aaboud$^\textrm{\scriptsize 34d}$,    
G.~Aad$^\textrm{\scriptsize 99}$,    
B.~Abbott$^\textrm{\scriptsize 124}$,    
O.~Abdinov$^\textrm{\scriptsize 13,*}$,    
B.~Abeloos$^\textrm{\scriptsize 128}$,    
D.K.~Abhayasinghe$^\textrm{\scriptsize 91}$,    
S.H.~Abidi$^\textrm{\scriptsize 164}$,    
O.S.~AbouZeid$^\textrm{\scriptsize 39}$,    
N.L.~Abraham$^\textrm{\scriptsize 153}$,    
H.~Abramowicz$^\textrm{\scriptsize 158}$
% ...
\bigskip
\\

% ...
$^{13}$Institute of Physics, Azerbaijan Academy of Sciences, Baku; Azerbaijan.\\
% ...
$^{34}$$^{(a)}$Facult\'e des Sciences Ain Chock, R\'eseau Universitaire de Physique des Hautes Energies - Universit\'e Hassan II, Casablanca;$^{(b)}$Centre National de l'Energie des Sciences Techniques Nucleaires (CNESTEN), Rabat;$^{(c)}$Facult\'e des Sciences Semlalia, Universit\'e Cadi Ayyad, LPHEA-Marrakech;$^{(d)}$Facult\'e des Sciences, Universit\'e Mohamed Premier and LPTPM, Oujda;$^{(e)}$Facult\'e des sciences, Universit\'e Mohammed V, Rabat; Morocco.\\
% ...
$^{39}$Niels Bohr Institute, University of Copenhagen, Copenhagen; Denmark.\\
% ...
$^{91}$Department of Physics, Royal Holloway University of London, Egham; United Kingdom.\\
% ...
$^{99}$CPPM, Aix-Marseille Universit\'e, CNRS/IN2P3, Marseille; France.\\
% ...
$^{124}$Homer L. Dodge Department of Physics and Astronomy, University of Oklahoma, Norman OK; United States of America.\\
% ...
$^{128}$LAL, Universit\'e Paris-Sud, CNRS/IN2P3, Universit\'e Paris-Saclay, Orsay; France.\\
% ...
$^{153}$Department of Physics and Astronomy, University of Sussex, Brighton; United Kingdom.\\
% ...
$^{158}$Raymond and Beverly Sackler School of Physics and Astronomy, Tel Aviv University, Tel Aviv; Israel.\\
% ...
$^{164}$Department of Physics, University of Toronto, Toronto ON; Canada.\\
% ...
$^{a}$ Also at Borough of Manhattan Community College, City University of New York, NY; United States of America.\\
$^{b}$ Also at Centre for High Performance Computing, CSIR Campus, Rosebank, Cape Town; South Africa.\\
$^{c}$ Also at CERN, Geneva; Switzerland.\\
$^{d}$ Also at CPPM, Aix-Marseille Universit\'e, CNRS/IN2P3, Marseille; France.\\
$^{e}$ Also at D\'epartement de Physique Nucl\'eaire et Corpusculaire, Universit\'e de Gen\`eve, Gen\`eve; Switzerland.\\
% ...
$^{*}$ Deceased

\end{flushleft}

% Created with Glance <Atlas.Glance@cern.ch>

\end{document}